\documentclass[twocolumn,showpacs,preprintnumbers,superscriptaddress,prb,floatfix,aps,10pt]{revtex4-1}

\usepackage{hyperref}
\hypersetup{
    bookmarks=true,         % show bookmarks bar?
    unicode=true,           % non-Latin characters in Acrobat’s bookmarks
    pdftoolbar=false,       % show Acrobat’s toolbar?
    pdfmenubar=true,        % show Acrobat’s menu?
    pdffitwindow=true,      % page fit to window when opened
    pdftitle={My title},    % title
    pdfauthor={Author},     % author
    pdfsubject={Subject},   % subject of the document
    pdfnewwindow=true,      % links in new window
    pdfkeywords={keywords}, % list of keywords
    colorlinks=true,        % false: boxed links; true: colored links
    linkcolor=blue,         % color of internal links
    citecolor=blue,         % color of links to bibliography
    filecolor=black,        % color of file links
    urlcolor=blue           % color of external links
  }

\usepackage{amsmath,amssymb}
\usepackage{graphicx,textcomp}
\usepackage{subfigure}
\usepackage{color}
\usepackage{todonotes}
\usepackage[titletoc,toc,title]{appendix}
\usepackage{xcolor}
\usepackage{soul}
\usepackage{todonotes}
\usepackage[utf8]{inputenc}

\definecolor{bjorn}{RGB}{240,226,182}
\definecolor{davide}{RGB}{211,22,112}

\begin{document}

\title{Longitudinal spin fluctuations in bcc and liquid Fe at high temperature and pressure calculated with a supercell approach}
\author{Davide Gambino}\email{davide.gambino@liu.se}
\affiliation{Department of Physics, Chemistry, and Biology (IFM), Link\"{o}ping University, SE-581 83, Link\"oping, Sweden.}

\author{Marian Arale Br\"{a}nnvall}
\affiliation{Department of Physics, Chemistry, and Biology (IFM), Link\"{o}ping University, SE-581 83, Link\"oping, Sweden.}

\author{Amanda Ehn}
\affiliation{Department of Physics, Chemistry, and Biology (IFM), Link\"{o}ping University, SE-581 83, Link\"oping, Sweden.}

\author{Ylva Hedstr\"{o}m}
\affiliation{Department of Physics, Chemistry, and Biology (IFM), Link\"{o}ping University, SE-581 83, Link\"oping, Sweden.}

\author{Bj\"{o}rn Alling}
\affiliation{Department of Physics, Chemistry, and Biology (IFM), Link\"{o}ping University, SE-581 83, Link\"oping, Sweden.}
%\affiliation{Max-Planck-Institut f\"ur Eisenforschung GmbH, D-402 37 D\"usseldorf, Germany.}

\date{\today}

\begin{abstract}

Investigation of magnetic materials at realistic conditions with first-principles methods is a challenging task due to the interplay of vibrational and magnetic degrees of freedom. 
The most difficult contribution to include in simulations is represented by the longitudinal magnetic degrees of freedom (LSF) due to their inherent many-body nature; nonetheless, schemes that enable to take into account this effect on a semiclassical level have been proposed and employed in the investigation of magnetic systems. However, assessment of the effect of vibrations on LSF is lacking in the literature. 
For this reason, in this work we develop a supercell approach within the framework of constrained density functional theory to calculate self-consistently the size of local-environment-dependent magnetic moments  in the paramagnetic, high-temperature state in presence of lattice vibrations and for liquid Fe in different conditions. 
First, we consider the case of bcc Fe at the Curie temperature and ambient pressure. 
Then, we perform a similar analysis on bcc Fe at Earth's inner core conditions, and we find that LSF stabilize non-zero moments which affect atomic forces and electronic density of states of the system. 
Finally, we employ the present scheme on liquid Fe at the melting point at ambient pressure, and at Earth's outer core conditions ($p \approx 200$ GPa, $T \approx 6000$ K). 
In both cases, we obtain local magnetic moments of sizes comparable to the solid-state counterparts.

\end{abstract}
\maketitle

\section{Introduction}

Magnetic materials find widespread application in many technological sectors, not only for their obvious magnetic properties \cite{PermanentMagnets} but also, for instance, as structural materials \cite{StructuralSteels,HighStrengthSteels}, best example of which are steels. Since modern society needs always more efficient devices, the design of materials has recently started to be guided by theoretical calculations \cite{TheoryGuidedExample_I,TheoryGuidedExample_II}. From this point of view, magnetism in solids poses a great challenge for modeling because of its many-body quantum nature and the difficulty in accurately representing thermal excitations, and for these reasons it has been and still is subject to thorough theoretical investigations\cite{StonerWolfarth,Moriya,DSFT}.

Solid state magnetism is historically described within two limits: the localized moments and the itinerant electron models \cite{COSSMS2016}. 
The former model, expressed in terms of a Heisenberg Hamiltonian, appropriately describes magnetic insulators, since the electrons responsible for magnetic effects are here localized in the atoms and give rise to robust magnetic moments, which interact with each other through different types of exchange mechanisms. 
The latter, instead, refers to magnetic metals, where the electrons are delocalized, and their cooperative interaction gives rise to magnetic properties; this type of systems has been usually investigated in terms of band ferromagnetism. 
These models are extremes of reality, and real systems fall in between these two classes.

A phenomenological theory that interpolates between the two limits was developed by Moriya \cite{Moriya}, which takes as fundamental variable the spin fluctuations. Its application to real systems was not widespread, although all following theories in this field are based on it. Recently, a dynamical version of Moryia's spin fluctuation theory has been established \cite{DSFT} and applied to simple systems like bcc Fe and Ni \cite{DSFTFeNi}, which seems to give quantitatively good results, although overestimating the Curie temperature.

To date, the only method that can account for magnetic effects in realistic systems of moderately large size is based on the Heisenberg model with parameters calculated from density functional theory (DFT). 
Theories that can take into account finite temperature quantum excitations in magnetic materials, such as dynamical mean field theory, are still confined to small system sizes\cite{DMFTcell}, rarely involving more than a unit cell \cite{DMFTvacancyFe}. 

The original Heisenberg model describes magnetic moments of constant size on a fixed lattice that interact with each other through well defined and constant exchange interactions. 
It can be employed in the investigation of more itinerant systems if one introduces more flexibility in the Hamiltonian: for a ferromagnetic metal, for instance, the exchange interactions result to be dependent on the local atomic and magnetic environment \cite{RubanSGPM,EisenbachFeJij,BjornJij,RubanFeJij}, therefore introduction of distance-dependent and magnetic state-dependent exchange interactions is needed in order to match experimentally available properties such as low temperature spin-wave excitations \cite{magnonsJij} or the Curie temperature \cite{RubanSGPM}.
One issue here is related to the modeling of the paramagnetic phase in DFT calculations, but this can be achieved with the use of the disordered local moment (DLM) approach in one of its different flavors \cite{CPADLM,SQSMSMDLM}. 
In particular, here we work in the framework of the magnetic sampling method (MSM) \cite{SQSMSMDLM}, in which many different random magnetic configurations are employed in the calculations, and properties of the paramagnetic state are then retrieved as statistical averages. Within this method, one can also include magnetic short range order effects to model finite temperature magnetism with the aid of constrained DFT calculations.

One further benefit of the MSM method is that it enables simulations which consider vibrational and magnetic degrees of freedom (DOFs) on a similar footing, a first example given by the disordered local moment molecular dynamics (DLM-MD) method \cite{DLMMD}. 
A further development of this concept, inspired from spin-lattice dynamics simulations, is the atomistic spin dynamics - \textit{ab} \textit{initio} molecular dynamics (ASD-AIMD) approach\cite{Irina_ASD-AIMD}, where interatomic forces are calculated within DFT, and the time evolution of the moments is propagated in parallel according to the Landau-Lifshitz-Gilbert (LLG) equation, with the two dynamics communicating to each other the positions of the atoms and the direction of the spins, which respectively determine the pair exchange interactions (and therefore the evolution of the moments) and the interatomic forces (and therefore the movement of the atoms). 

In this type of simulations, only transversal spin fluctuations have been included so far\cite{Irina_ASD-AIMD}, neglecting longitudinal spin fluctuations (LSF); these are excitations along the direction of the magnetic moment that are not directly included in the standard Heisenberg Hamiltonian, and they are more intimately connected with the quantum nature of electrons as compared to transversal fluctuations. 
Regular DFT calculations do not reproduce this effect because of the independent-particle nature of the Kohn-Sham electrons. 
Nonetheless, semiclassical models to include LSF on a mean-field basis have been developed in the past years \cite{Moriya,LSFMurata,LSFKubler,LSFRosengaard,LSFRuban,LSFSandratskii,classicalLSFWysocki,LSFVitos,LSFPan,LSFKhmelevskyi}.

From the work of Murata and Doniach\cite{LSFMurata}, in which a fourth-order dependence of the energy functional on the magnetic moment, inspired from the Ginzburg-Landau expansion, was introduced, many applications of this semiclassical approach have been performed on more or less itinerant systems. 
Uhl and K\"ubler \cite{LSFKubler} have applied this concept to bcc Fe, fcc Ni, and fcc and hcp Co by calculating all parameters of the model Hamiltonian with \textit{ab} \textit{initio} calculations of spin-spirals, considering fluctuations in reciprocal space. Similarly, Rosengaard and Johansson \cite{LSFRosengaard} investigated finite temperature magnetism of the same systems but with a real-space formulation of the problem, and employing the calculated parameters in Monte Carlo (MC) simulations to derive properties at and above the Curie temperature; for bcc Fe, they find a weakly increasing local magnetic moment as a function of temperature.
Ruban \textit{et} \textit{al.} \cite{LSFRuban} investigated LSF in a DLM-CPA framework with a slight modification of the Heisenberg-Landau Hamiltonian, and performed MC simulations based on these parameters calculated from first-principles, finding a decrease in size of the average magnetic moment for bcc Fe.
Wysocki \textit{et} \textit{al.} \cite{classicalLSFWysocki} studied more in general the thermodynamics of itinerant magnets in a  classical spin fluctuation model, focusing on the problems related to phase space measure (PSM).
More recently, analytical expressions for the magnetic entropy have been introduced in the single-particle potentials of DFT calculations in order to approximately include LSF in calculations\cite{RubanEarthCore,KorzhavyiFe}.
With this method, Dong \textit{et} \textit{al.} \cite{LSFVitos} have calculated the thermal expansion in Fe based on a Debye-Gr\"uneisen model.

Despite the rich literature, no direct investigation of the interplay between lattice vibrations and LSF can be found.
For this reason, in this work we develop a method to calculate the on-site energy landscape that governs LSF and self-consistently calculate the magnetic moment at a given temperature for each atom in a vibrating lattice. 
Since LSF are more important in the paramagnetic state, we focus only on this regime.
We test the method on a snapshot taken from an ASD-AIMD simulation of bcc Fe at its Curie temperature, and on bcc Fe at Earth's inner core conditions employing a snapshot taken from a nonmagnetic (NM) MD run \cite{SergeiFeHTHP}. 
In addition, we investigate also liquid Fe at ambient pressure and around the melting point, and at conditions expected in the Earth's outer core ($p\approx 200$ GPa, $T\approx 6000$ K), also in this case taking a single snapshot from MD simulations.
We observe that the local environment strongly affects the shape of the on-site energy landscape, and the inclusion of LSF in the calculation affects pressure and forces in the considered snapshot.

The article is structured as follows: in Sec. \ref{SectionMethods} the semiclassical theory of LSF is reviewed (Sec. \ref{SectionLSFtheory}), with special attention to the issue of PSM, and the scheme for the calculation of the on-site energy and determination of the size of magnetic moments at a given temperature in a vibrating lattice is presented (Sec. \ref{SectionSupercellApproach}), with computational details and the origin of the snapshot presented in Sec. \ref{SectionComputationalDetails}. 
In Sec. \ref{SectionResultsbccFeAP} the scheme is tested on bcc Fe at ambient pressure and temperature close to the Curie temperature, and the effect of vibrations and magnetic disorder is inferred, together with an evaluation of the effect of different PSM on the size of the magnetic moments.
The application of the method on bcc Fe at Earth's inner core conditions is then presented in Sec. \ref{SectionResultsbccFeHTHP} and results are compared with previous investigations. Sec. \ref{SectionResultsLFe} deals with liquid Fe at the melting point under ambient pressure conditions, and liquid Fe at outer Earth's core conditions.
Finally, in Sec. \ref{Conclusions} the conclusions of the work are summarized.

\section{Theoretical methods} \label{SectionMethods}

\subsection{Semiclassical theory of longitudinal spin fluctuations} \label{SectionLSFtheory}

For a system of $N$ magnetic moments, a generalized Heisenberg Hamiltonian can be written as:

\begin{equation} \label{EqGHH}
H=-\sum_{i\neq j} J_{ij} \textbf{m}_i \cdot \textbf{m}_j + \sum_i E_i(m_i),    %(\textbf{R}_i,\textbf{R}_j)
\end{equation}
where $J_{ij}$ is the exchange interaction between moment $\textbf{m}_i$ and moment $\textbf{m}_j$, and $E_i(m_i)$ is the on-site term that depends on the size of the moment $m_i$; bold symbols represent vectors here. 

In general, both $J_{ij}$ and $E_i(m_i)$ can depend on the local arrangement of the atoms and the relative magnetic moments, with possible variations in these quantities as a function of the surrounding environment, depending on the nature of the system under investigation. 
The last term in Eq. \ref{EqGHH} represents the energy of moment $\textbf{m}_i$, immersed in the mean-field created by all the other magnetic moments, as a function of its own magnitude: this is the semiclassical term associated to LSF. 
For a system with very well localized magnetic moments, this term does not depend on the surrounding moments and has typically a sharp minimum, but for a more itinerant system the magnetic state plays a crucial role \cite{LSFRuban}. Its general form is inspired from Ginzburg-Landau thermodynamic theory of magnetism:

\begin{equation}\label{EqEnergyPolynomial}
E_i(m_i)=\sum_{n=0}^{\infty} a_n m_i^{2n}\approx a m_i^2+b m_i^4,
\end{equation}
where at least a 4th order polynomial is needed to approximate a localized moment landscape.

Thermodynamic quantities for the Hamiltonian in Eq. \ref{EqGHH} can be derived from the partition function:

\begin{equation}
Z=\int d\textbf{m}_1 d\textbf{m}_2 \ldots d\textbf{m}_N \; e^{-\frac{H}{k_B T}},
\end{equation}
with $k_B$ being the Boltzmann constant, $T$ the temperature, and the integration extends to all possible configurations of the moments $\textbf{m}_i$. In the paramagnetic state, the first term on the right-hand side of Eq. \ref{EqGHH} is equal to zero; therefore, since the on-site term does not depend explicitly on other moments, the partition function can be written as the product of $N$ partition functions $Z=\prod Z_i$, one for each moment $\textbf{m}_i$, with single-moment partition function $Z_i$ defined as:	

\begin{equation}
Z_i=\int d\textbf{m}_i \; e^{-\frac{E_i(m_i)}{k_B T}}=\int_0^{\infty} dm_i \; \textrm{PSM} \; e^{-\frac{E_i(m_i)}{k_B T}}.
\end{equation}
Here, the last expression is obtained passing to spherical coordinates and integrating out the angular ones, since the energy $E_i$ does not depend explicitly on them in the paramagnetic state in this approximation. 
Constants have been dropped because they cancel out or add an irrelevant constant shift to thermodynamic quantities.
PSM is the phase space measure, which from the above derivation is apparently just the Jacobian $m^2$; however, this term is not well defined and it will be more thoroughly discussed later in this section.

Thermodynamic quantities can be calculated from here using well-known relations. As an example, the average size of the moment at temperature $T$, $\langle m_i (T)\rangle$, for a given energy landscape $E_i$ is calculated as:

\begin{equation}\label{EqThermodynamicAvM}
\langle m_i (T) \rangle = \frac{1}{Z_i}\int_0^\infty dm_i \; \textrm{PSM} \; m_i \; e^{-\frac{E_i(m_i)}{k_B T}}.
\end{equation}

The partition function of a magnetic moment in a system in the paramagnetic state can be analytically calculated only for a quadratic form of the energy $E_i$ as a function of the moment size $m_i$; in this case, one can also derive an analytical expression for the magnetic entropy as a function of the average moment \cite{RubanEarthCore}. In the case of $\textrm{PSM}=m^2$ and considering a system in the paramagnetic state, this is $S^{\textrm{mag}}=3k_b\log\langle m \rangle$, which differs from the entropy derived from the quantum Heisenberg model $S^{\textrm{mag}}=k_b\log(m+1)$ where LSF are not considered. 
Schemes to include these analytical expressions of the magnetic entropy directly in the one-electron potentials in DFT calculations have also been developed \cite{RubanEarthCore,RubanFeJij,KorzhavyiFe}.
However, a harmonic expression for the energy landscape is often very approximate; in fact, a fourth-order polynomial is more suited in both partly localized systems (such as bcc Fe, where this S$^{\textrm{mag}}$ would overestimate the magnetic moment) and itinerant ones (as, e.g., Ni, where the moment is underestimated \cite{RubanInvar}).
Therefore, numerical integration of the partition function needs to be performed in order to obtain thermodynamic properties of the system.

\begin{figure}
\begin{center}
\includegraphics[width=0.5\textwidth]{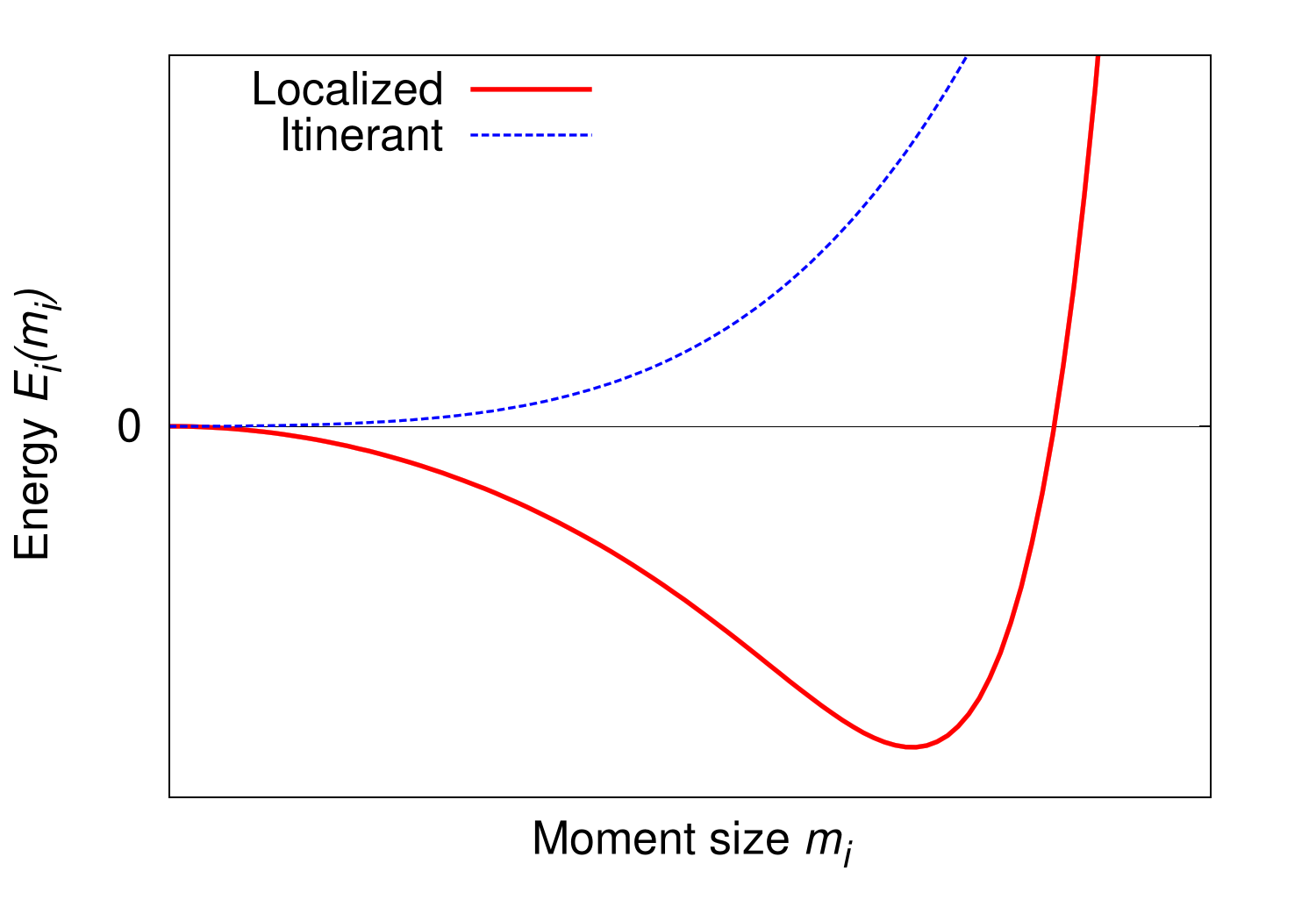}
\caption{Example of the typical energy landscapes for a localized moment (solid line) and an itinerant (dashed line) system. \label{FigLocalizedVSItinerant}}
\end{center}
\end{figure}

At this point, the issue of PSM should be addressed. First of all, different types of systems (localized moments or itinerant electrons) show different energy landscapes (see Fig. \ref{FigLocalizedVSItinerant}), with the localized moment system having a minimum at $m\neq 0$, whereas the itinerant one has minimum for $m=0$.
Local moments in the itinerant system can be present only because of temperature.
In general, at finite temperatures, the moments can evolve in time by rigid rotations (transversal fluctuations) and/or by changing their size (longitudinal fluctuations), where the former evolution is slower than the latter; the predominant mechanism is dictated by the energy landscape $E(m)$. 
For an itinerant system, assuming that at time $t=0$ a moment has value $m=\bar{m}$, the moment can change direction by shrinking down to 0 and then reappearing in another direction. 
This evolution mechanism is fast as compared to the rigid rotation of the moment, therefore coupling longitudinal and transversal DOFs. 
For this reason, the full dimensionality of the problem has to be taken into account when performing thermodynamic averages, and from here the use of $\textrm{PSM}=m^2$ is motivated. 
On the opposite, for a localized moment system, the predominant mechanism of evolution is the rotation of the finite moment, therefore the longitudinal fluctuations can be treated as adiabatically fast and confined to the direction of the moment, motivating the use of a $\textrm{PSM}=1$ as required for a monodimensional problem.
Intermediate PSMs have also been suggested \cite{LSFKhmelevskyi}, however no rigorous method to define which PSM should be used in which case is available.
In this work, we do not aim at solving the problem of PSM, but we think that the reader should be aware of it, and we show the consequences of different choices of PSM in Sec. \ref{SectionResultsbccFeAP}.

\subsection{Supercell scheme for calculation of magnetic moments size at finite temperature} \label{SectionSupercellApproach}

The calculation of the energy landscapes as the ones shown in Fig. \ref{FigLocalizedVSItinerant} is easily performed with the aid of constrained DFT calculations, and it can be carried out in both magnetically ordered and disordered states \cite{LSFRuban}. In this article, we devise a scheme to calculate the landscapes at a given temperature which does not rely on symmetry in the system, therefore applicable to a vibrating lattice or a system with, e.g., topological disorder like liquids.

\begin{figure}
\begin{center}
\includegraphics[width=0.5\textwidth]{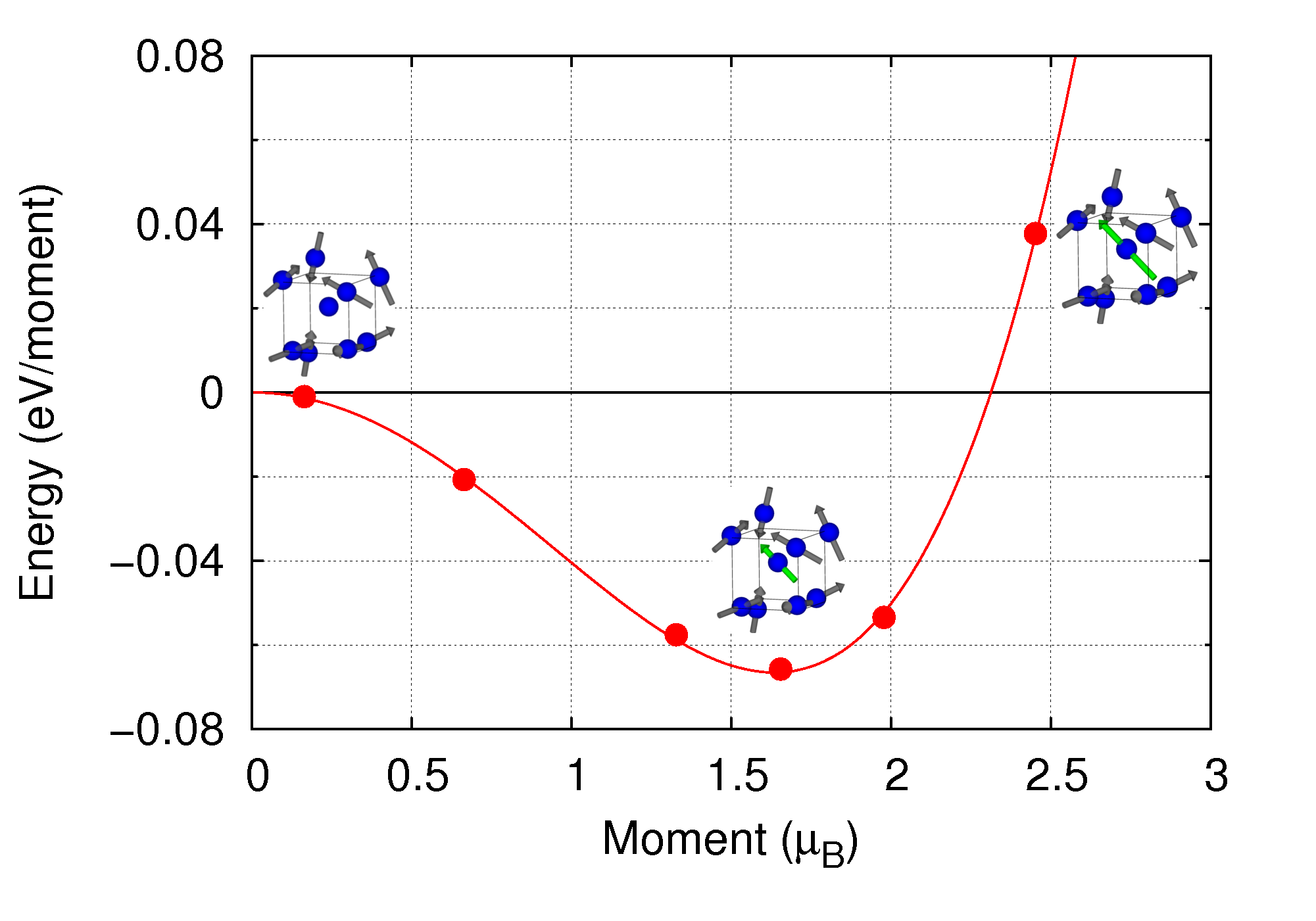}
\caption{Illustration of the procedure of the calculation of the energy landscape, where the energy points in the graph are calculated for different sizes of the moment on the central atom in the cartoons, while keeping fixed all the others. \label{FigSchemeCalculation}}
\end{center}
\end{figure}

Given N atoms which, for simplicity, will be considered all magnetic, then a particular atomic-magnetic configuration is defined by the position of all atoms $\{\textbf{R}_1,...,\textbf{R}_N\}$ and the direction and size of the moments associated to each of them $\{\textbf{m}_1,...,\textbf{m}_N\}$ (for the source of this atomic-magnetic configuration, the reader is referred to Sec. \ref{SectionComputationalDetails}).
As an example, if the system has at least partly localized moments, a DFT calculation constraining only the direction of the moments along some pre-established direction will give the moments size that correspond to ``0 K'' temperature on the longitudinal DOFs, whereas for a system with ``0 K'' nonmagnetic solution (itinerant), then the initial magnetic configuration would be with zero moments. 
The moments do not need to have all the same magnitude even if the system involves only one element, since we allow for disorder in atomic positions and moments direction, therefore every moment experiences a unique environment.

Once the ``0 K'' size of the moments is established, we proceed to calculate the energy landscape as a function of moment size for each atom in the supercell, which again can be different from each other due to the different local environment. 
To do this for a specific atom $i$, we fix the size and direction of all the other moments to the ``0 K'' value previously calculated, and we perform completely constrained calculations (i.e., constraining both direction and magnitude of the moments) varying only the size of moment $m_i$ in steps. We generally employ six different magnitudes of each moment $m_i$.
Then, the specific energy landscape $E_i(m_i)$ is obtained by fitting a fourth order polynomial as in Eq. \ref{EqEnergyPolynomial} to the results of the constrained calculations for different values of $m_i$.
An example of the energy vs moment size calculated in this way for moment $\textbf{m}_i$ and the polynomial fit are shown in Fig. \ref{FigSchemeCalculation} .

This procedure is repeated for all atoms in the supercell, so that an energy landscape is obtained for each atom. The thermodynamic average moment of atom $i$ at a certain temperature $\bar{T}$, $\langle m_i(\bar{T}) \rangle $, calculated here numerically with Eq. \ref{EqThermodynamicAvM}, is chosen as the moment of atom $i$ at $T=\bar{T}$. This average is just the pure thermodynamic average from Eq. \ref{EqThermodynamicAvM}, which does not involve an average over atoms in the supercell.

Since at this point the size of the moments at temperature $\bar{T}$ is different from the ``0 K'' value, the energy landscapes $E_i$ might have changed. Therefore, the same procedure is reiterated for all the moments until $\langle m_i(\bar{T}) \rangle$ does not considerably change as compared to the previous iteration for each single atom, with a convergence criterion of $0.01 \mu_B$.

\subsection{Computational details and origin of the atomic configurations} \label{SectionComputationalDetails}

All calculations are carried out with the Vienna ab initio simulation package VASP \cite{VASP_I,*VASP_II,*VASP_III,*VASP_IV} employing projector-augmented wave (PAW) potentials \cite{PAW_Blochl,PAW_vasp}. 
54 atoms supercells (3x3x3 conventional bcc unit cells) are employed in the case of Fe at ambient pressure and $T\approx T_C=1043$ K, and bcc Fe at Earth's inner core conditions ($T\approx 6000$ K, $p\approx 300$ GPa). 
For the liquid cases, we consider cubic boxes with 54 and 125 atoms for the high-pressure and the ambient pressure cases, respectively.
In the high-pressure cases, the $3s$ and $3p$ electrons are considered as valence electrons. All calculations are performed with noncollinear magnetism.
 
For bcc Fe at ambient pressure, the atomic-magnetic configuration employed in the investigation is a snapshot of an ASD-AIMD run, performed with Langevin dynamics to control the temperature (damping parameter 1 ps$^{-1}$), cutoff energy for expansion of plane waves 400 eV, and the first Brillouin zone sampled with a $\Gamma$-centered 2x2x2 mesh for the AIMD part, evolving the atomic trajectories with a 1 fs timestep. 
The direction of the moments are constrained with the method presented in Ref. [\onlinecite{MaDudarevCDFT}], employing a constraining parameter $\lambda=10$. 
For what concerns the ASD part, which is performed with the UppASD code \cite{UppASD_I,UppASD_II}, we employ a 0.01 fs timestep for evolution of the moments direction and a damping parameter of 0.05 in the LLG equations. After equilibration, a combined atomic-magnetic configuration is taken for investigation of LSF. 
As previously mentioned, the ``0 K'' moments are here obtained from a DFT calculation constraining only the direction of the moments.
In this particular case, the direction of the moments is the one derived directly from the ASD-AIMD run, since within this approach magnetic and atomic configurations are coupled.

Regarding bcc Fe at Earth's inner core conditions, the snapshot is taken from a NM MD run, the results of which are reported in Ref. [\onlinecite{SergeiFeHTHP}]. For the computational details regarding this simulation, the reader is referred to the original paper. In the calculation of LSF in this case, we employ moments with random directions, with first iteration performed for each atom in a NM background.

For ambient pressure liquid Fe, where the temperature chosen is around the melting point ($ T_m= 1823 $ K), the positions employed are obtained from a DLM-MD simulation\cite{DLMMD}.
More in detail, the simulation was started by placing 125 Fe atoms in a cubic box at random positions matching the experimental density at T$_m$. 
A NM run was then performed for 500 fs, and from these positions a DLM-MD simulation was carried out with a timestep of 1 fs and changing the collinear magnetic configuration every 5 fs. 
The total length of the simulation was about 6.5 ps. Fermi-Dirac smearing was included in the calculations, with electronic temperature matching the melting temperature.
The atomic configuration for the LSF analysis was randomly picked from the DLM-MD run. The details of the magnetic configuration are explained in Sec. \ref{SectionResultsLFe}.

The high-pressure liquid case, similarly to high-pressure bcc Fe, is investigated from a snapshot of a NM MD simulation at $T=6000$ K and $p\approx 200$ GPa, conditions expected in Earth's outer core.

The DFT calculations of the energy landscapes are performed with a thicker k-points mesh (5x5x5 Monkhorst-Pack mesh \cite{MPscheme}), and constraining both direction and size of the moments with the method implemented in VASP with $\lambda=25$.

\section{Results}
\subsection{Bcc Fe at the Curie temperature} \label{SectionResultsbccFeAP}

\begin{figure}[t!]
\begin{center}
\includegraphics[width=0.5\textwidth]{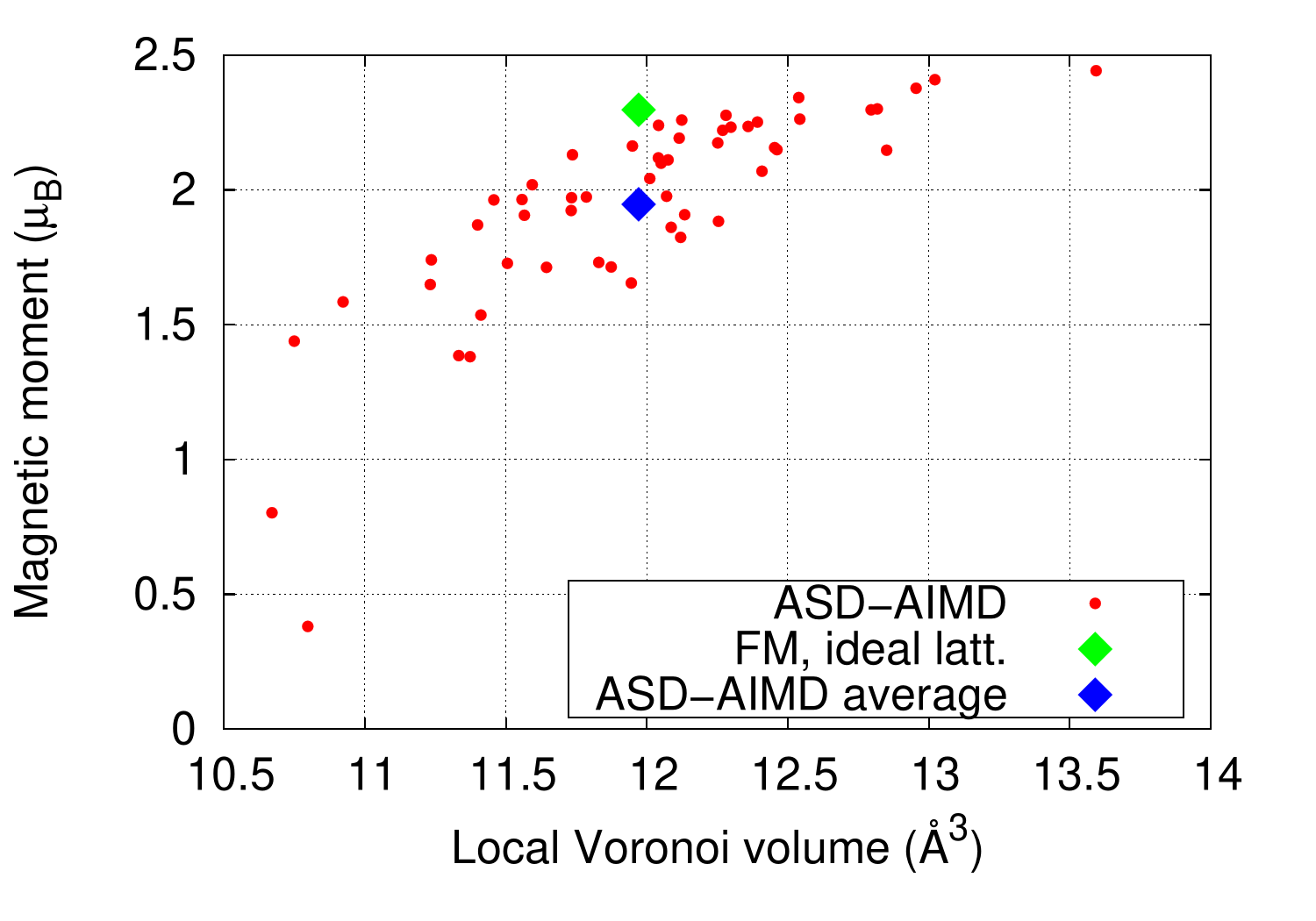}
\caption{Magnetic moments vs local Voronoi volume for bcc Fe from the ASD-AIMD simulation at the Curie temperature; a single value for each moment is obtained here because only the direction of the moments is constrained. \label{FigVorvolMag}}
\end{center}
\end{figure}

As a starting point of the analysis, we calculate the ``0 K'' size of the magnetic moments, shown in Fig. \ref{FigVorvolMag} as a function of local Voronoi volume.
As it can be seen, the moment size shows a certain degree of correlation with the local Voronoi volume, with smaller moments for smaller volumes. The Voronoi volume and moment size for an ideal FM bcc lattice and the average moment for the present configuration are also shown as a comparison (green and blue diamond, respectively). We observe that, as already known\cite{BjornFe}, the average moment size is lower in this  particular configuration as compared to the FM moments on ideal lattice.

\begin{figure}[t!]
\begin{center}
\includegraphics[width=0.5\textwidth]{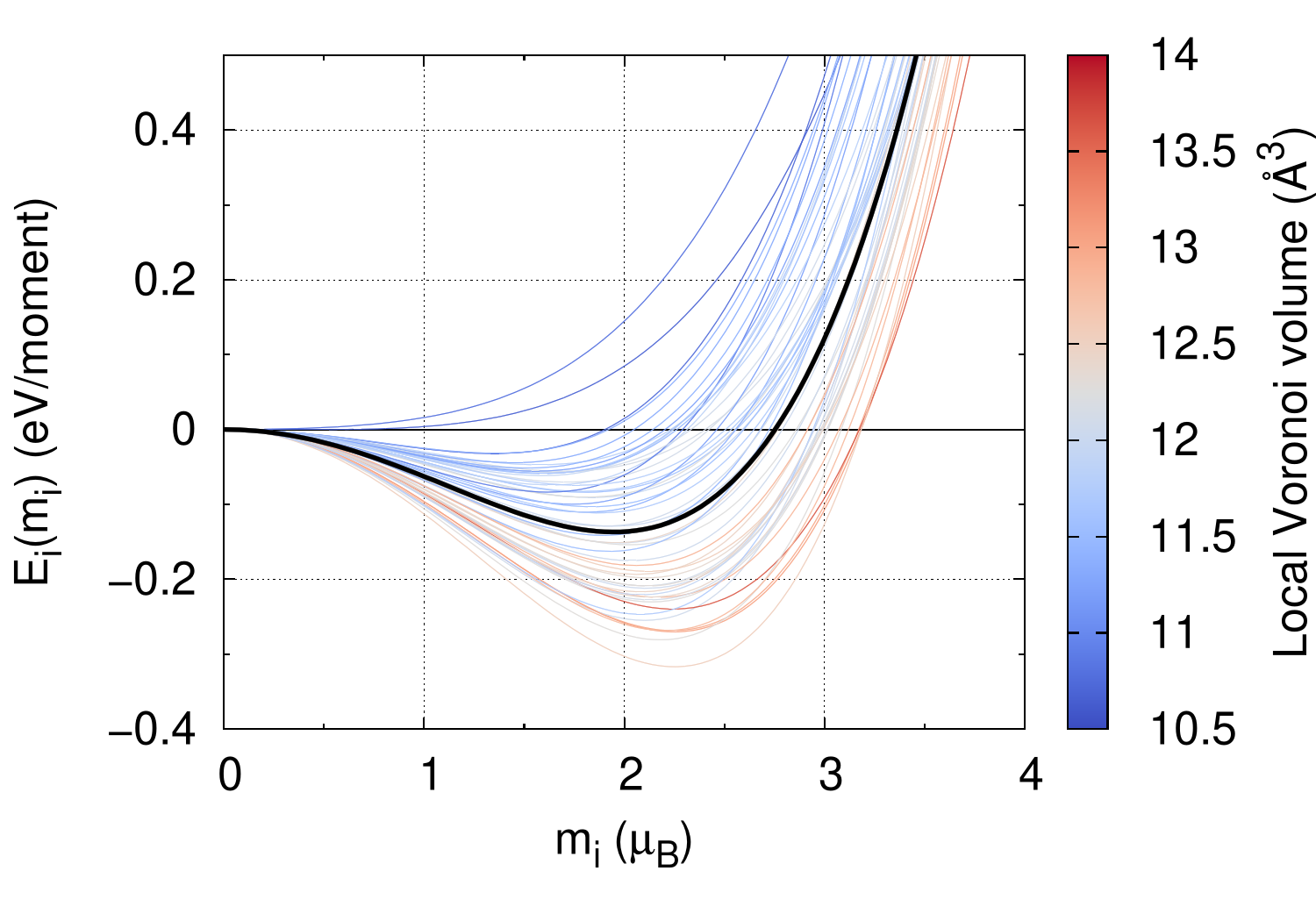}
\caption{Energy landscapes calculated with the present method for each atom in bcc Fe at the Curie temperature from the ASD-AIMD snapshot with $\textrm{PSM}=m^2$. The color code indicates the size of the local Voronoi volume for the given atom. The black landscape indicates the average. \label{FigLandscapes}}
\end{center}
\end{figure}

At this point, the calculation of the energy landscapes for each moment $m_i$ is carried out according to the procedure described in Sec. \ref{SectionSupercellApproach}. We calculate the effect of LSF at $T=1100$ K and we employ both the monodimensional and three-dimensional PSM. In both cases, we reach convergence after three iterations. 
The fitted energy landscapes are shown in Fig. \ref{FigLandscapes} for $\textrm{PSM}=m^2$; use of the other PSM gives differences in the landscapes, but difficult to discern by eye.

As it can be seen, the local environment affects strongly the shape of these moments, with two cases in which the landscapes are more similar to an itinerant rather than localized moment system, as bcc Fe is usually considered. 
These moments belong to atoms characterized by particularly small local Voronoi volumes.
Of course, this is a special case; on average during a simulation, one can expect to recover a more localized moment behavior. 
However, the dependence on the local environment affects the instantaneous forces, as it will be shown later in this section, and therefore the dynamics. These effects might play an important role in the proper sampling of the phase space in, e.g., an AIMD simulation.

A further detail that should be noticed concerns the two moments with a more itinerant-like behavior. As seen in Fig.  \ref{FigVorvolMag}, no moment has actually a zero value in the initial calculation with only direction constrained, however in the fully constrained calculations these two moments have minimum energy for $m=0$. 
This is probably due to the full constraint itself, i.e., when we constrain only the direction of the moments, the system can rearrange the electrons to minimize the energy leading to nonzero moments, whereas with full constraint there is less room for rearrangement  and the lowest energy solution has now $m=0$. Another reason for this discrepancy could be that the direct ``0 K'' calculation gets stuck in a local minimum before reaching $m=0$ because of numerics of the DFT code, which could happen for a very flat energy landscape.

\begin{figure}[t!]
\begin{center}
\includegraphics[width=0.5\textwidth]{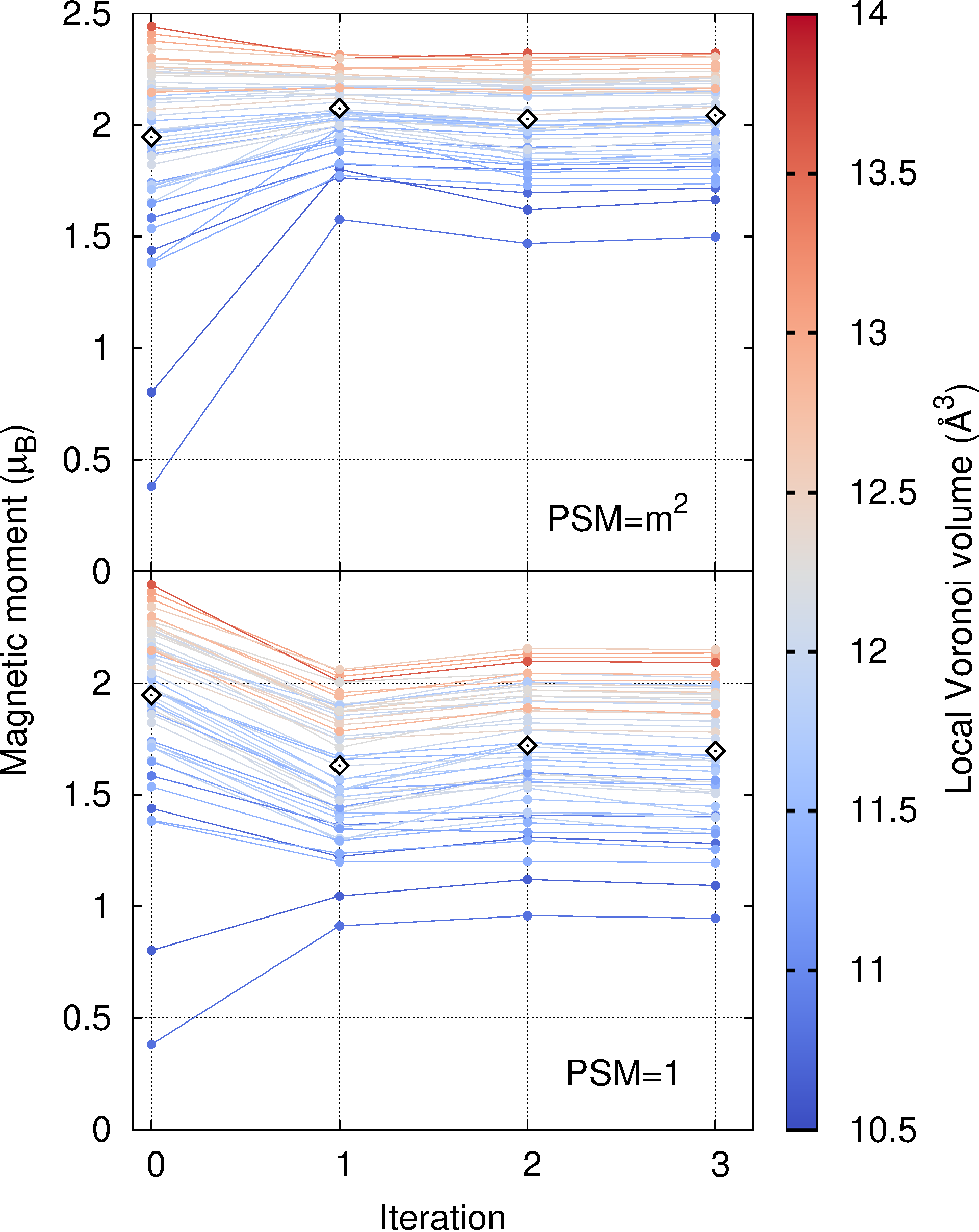}
\caption{Moment size with inclusion of LSF at each iteration of the procedure for $\textrm{PSM}=m^2$ (top) and $\textrm{PSM}=1$ (bottom). The color of the lines and symbols is related to the size of the local Voronoi volume of the atom, whereas the empty diamonds are the average moment at each iteration. \label{FigMomLSF}}
\end{center}
\end{figure}

The size of the moments with inclusion of LSF at temperature $T=1100$ K at each iteration is shown in Fig. \ref{FigMomLSF} for the different PSMs, and they are compared to the ``0 K'' value (the $0^{\textrm{th}}$ iteration). 
Between the first and the second iteration, the moment size is generally already converged within $\approx 0.2\; \mu_B$ for $\textrm{PSM}=m^2$, whereas for the monodimensional PSM the convergence is slower due to the larger differences in the size of most of the moments with LSF. 
In addition, it is interesting to notice that in the former case, LSF induce an increment in the average moment size (black empty diamonds in Fig. \ref{FigMomLSF}), whereas in the latter this quantity decreases.

The spread in landscapes due to local environments seems to suggest that bcc Fe cannot be treated as a fully localized moment system, therefore $\textrm{PSM}=1$ is probably not the relevant PSM to consider. 
A certain degree of coupling between longitudinal and transversal DOFs would lead to results intermediate between the present ones. 
An additional comment regarding $\textrm{PSM}=1$ concerns the limits of integration in the partition function: since in this case the problem is monodimensional, the integration space should be extended to negative values of the moments. In the present case of an even-power landscape, no difference would be found in average moment size; on the other hand, usage of a polynomial which includes odd-powers would require fitting and integration in the negative-moment space. In any case, within the assumptions of the present study, this is not an issue.
Finally, when using $\textrm{PSM}=1$, one needs to include contribution to the entropy from the transverse DOFs outside the LSF scheme, e.g. by consideration of a Heisenberg model.

\begin{figure}[t!]
\begin{center}
\includegraphics[width=0.5\textwidth]{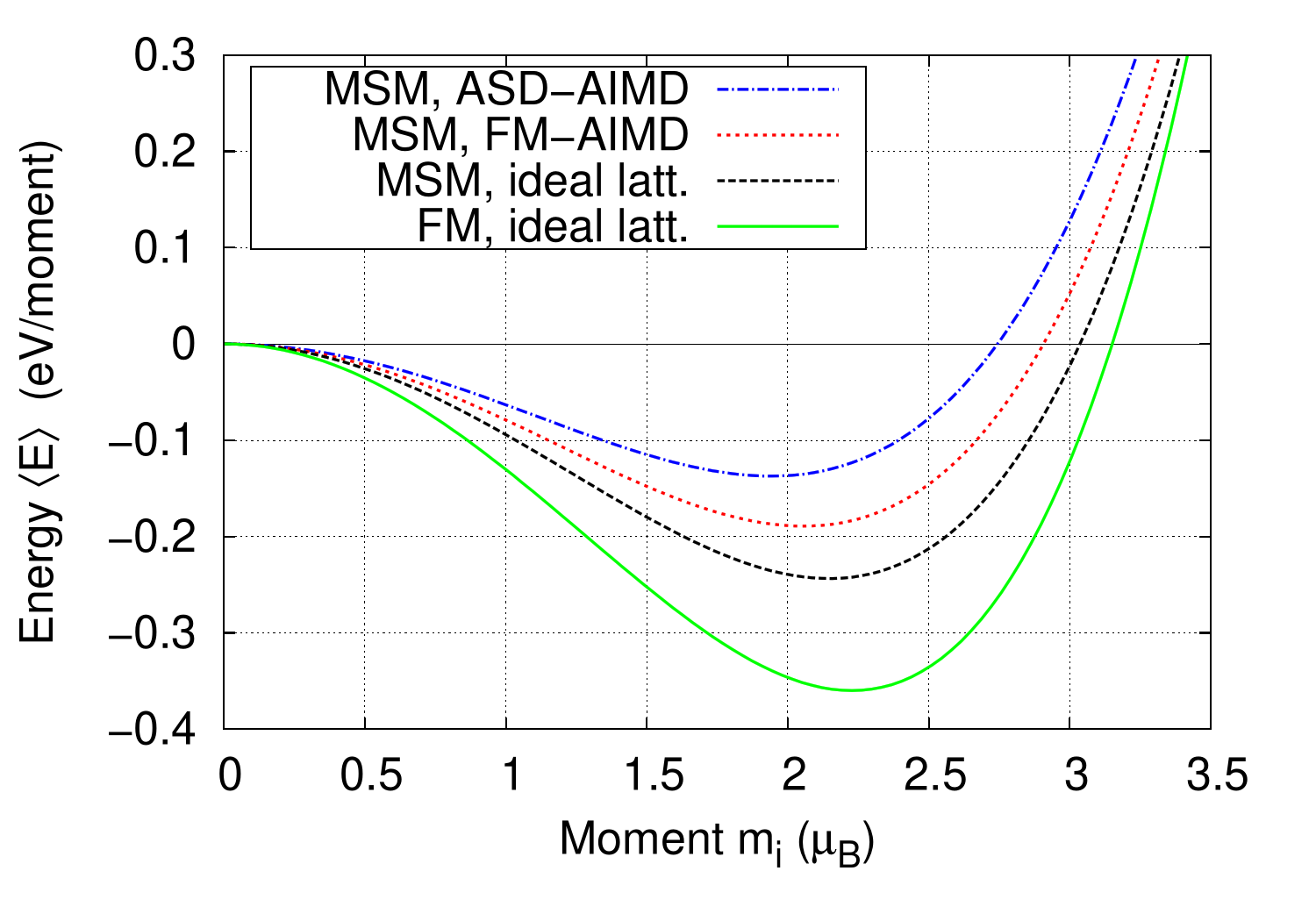}
\caption{Effect of different degrees of vibrational and magnetic disorder on the average energy landscape in bcc Fe. The magnetic sampling method (MSM) models the paramagnetic state. \label{avLandscapeDisorderEffect}}
\end{center}
\end{figure}

In order to estimate the effect of lattice and magnetic disorder on the energy landscapes, we perform the calculation of $E_i(m_i)$ with different degrees of disorder, starting from FM moments on ideal bcc lattice, then the same instantaneous magnetic configuration (referred to as ``MSM'') obtained from the ASD-AIMD simulation but on the ideal lattice, MSM on atomic positions obtained from a FM-AIMD simulation, and finally the average landscape from Fig. \ref{FigLandscapes}. 
For these landscapes, we perform only one iteration of the procedure, since already at this level of convergence conclusions on the effect of disorder can be inferred. 
In Fig. \ref{avLandscapeDisorderEffect}, the average landscapes for the different atomic-magnetic configurations are presented. 
It can be immediately seen that, for increasing degree of disorder, the average landscape becomes shallower and shallower, and the position of the minimum shifts towards lower magnetic moment sizes. 
It is worth to mention again that these landscapes are given for a particular atomic-magnetic configuration with 54 atoms, therefore if the procedure would be performed on more configurations, a change in the quantitative values of the average landscapes could be expected; nonetheless, we expect the qualitative picture to remain the same.

\begin{figure}[t!]
\subfigure{\label{FigFmodbccFeHTRP} \includegraphics[width=0.5\textwidth]{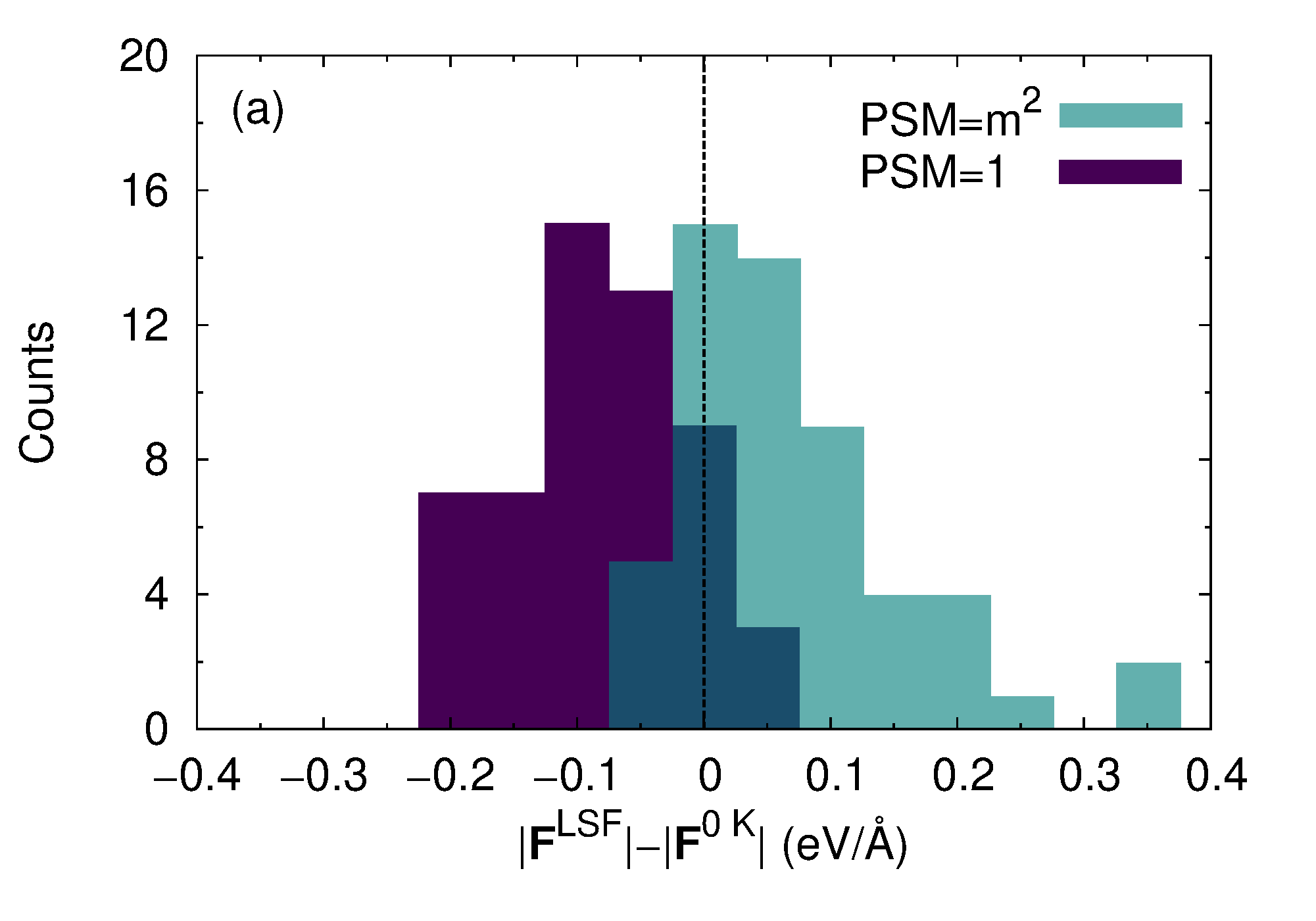}}
\subfigure{\label{FigAngFbccFeHTRP} \includegraphics[width=0.5\textwidth]{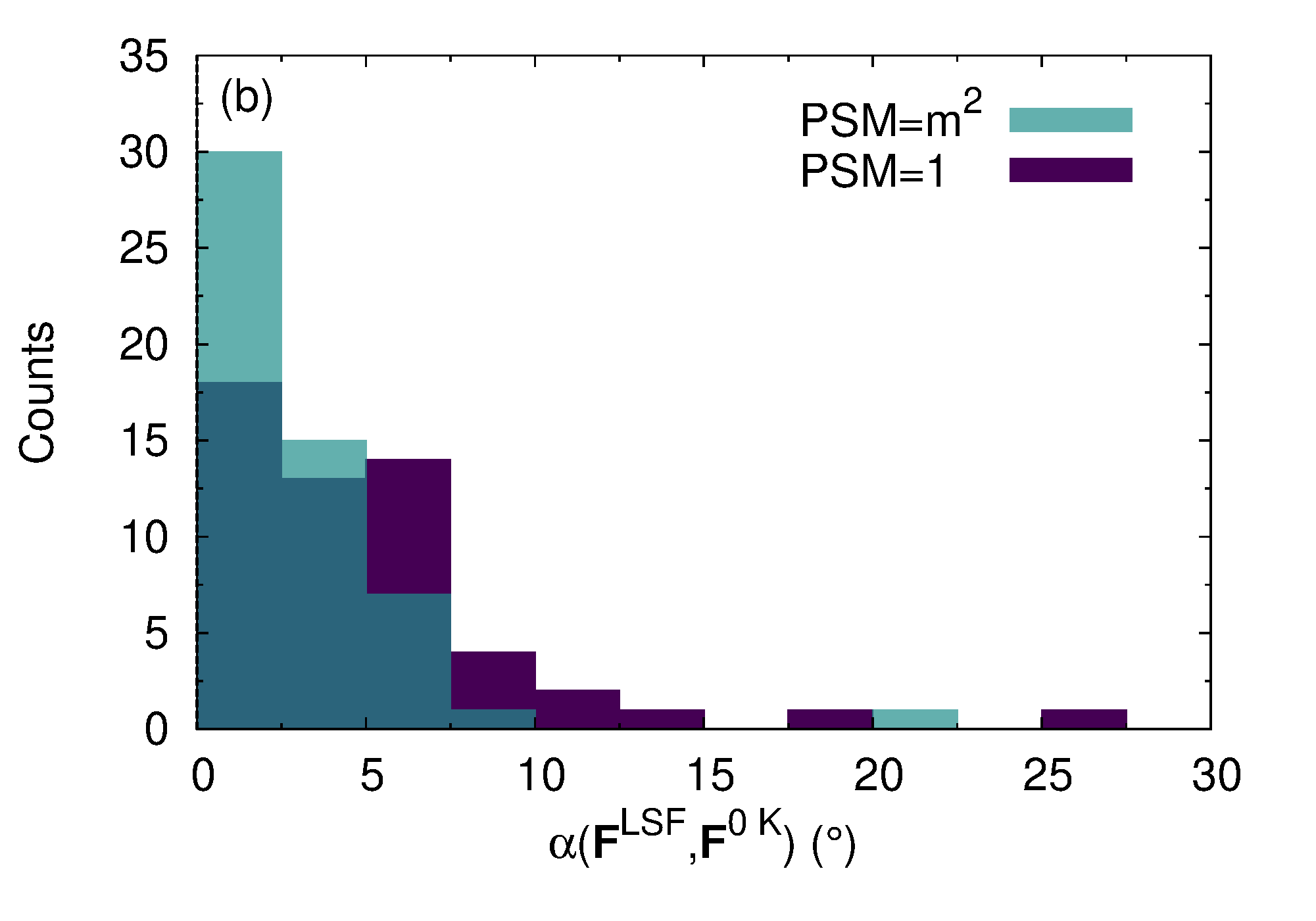}}
\caption{(a) Difference in modulus of the interatomic forces calculated with different PSMs and the ``0 K'' value. (b) Angle between forces with and without inclusion of LSF.  \label{FigForcesbccFeHTRP}}
\end{figure}

In this context, it is important to evaluate the effect of LSF on the forces acting on the atoms, which will affect the evolution of the atomic positions during an AIMD simulation, therefore affecting the vibrational phase-space sampling.
The change in intensity and direction of the forces with LSF for the two different PSMs as compared to the ``0 K'' moments results is depicted in Fig. \ref{FigForcesbccFeHTRP}. 
Forces obtained with $\textrm{PSM}=m^2$ tend to be larger than the ``0 K'' forces, due to the general increase in magnetic moment size; for $\textrm{PSM}=1$, the opposite happens (Fig. \ref{FigFmodbccFeHTRP}). The change in forces modulus and direction is in general relatively small ($\approx 0.1$ eV/\AA, $0-5 ^{\circ}$), however for some atoms deviations are more important. Inclusion of LSF affects also the pressure in the supercell, increasing of $\approx 1.5$ GPa with $\textrm{PSM}=m^2$, and decreasing of $\approx 5$ GPa with $\textrm{PSM}=1$. The lattice parameter employed here is the 0 K theoretical lattice parameter expanded with the experimental thermal expansion coefficient. 
Since the ``0 K'' atomic-magnetic configuration is at a pressure of $\approx -5$ GPa, the inclusion of LSF with $\textrm{PSM}=m^2$ seems to be an improvement.

\subsection{Bcc Fe at Earth's inner core conditions} \label{SectionResultsbccFeHTHP}

\begin{figure}[t!]
\subfigure{ \label{FigLandscapesFeHTHP}\includegraphics[width=0.5\textwidth]{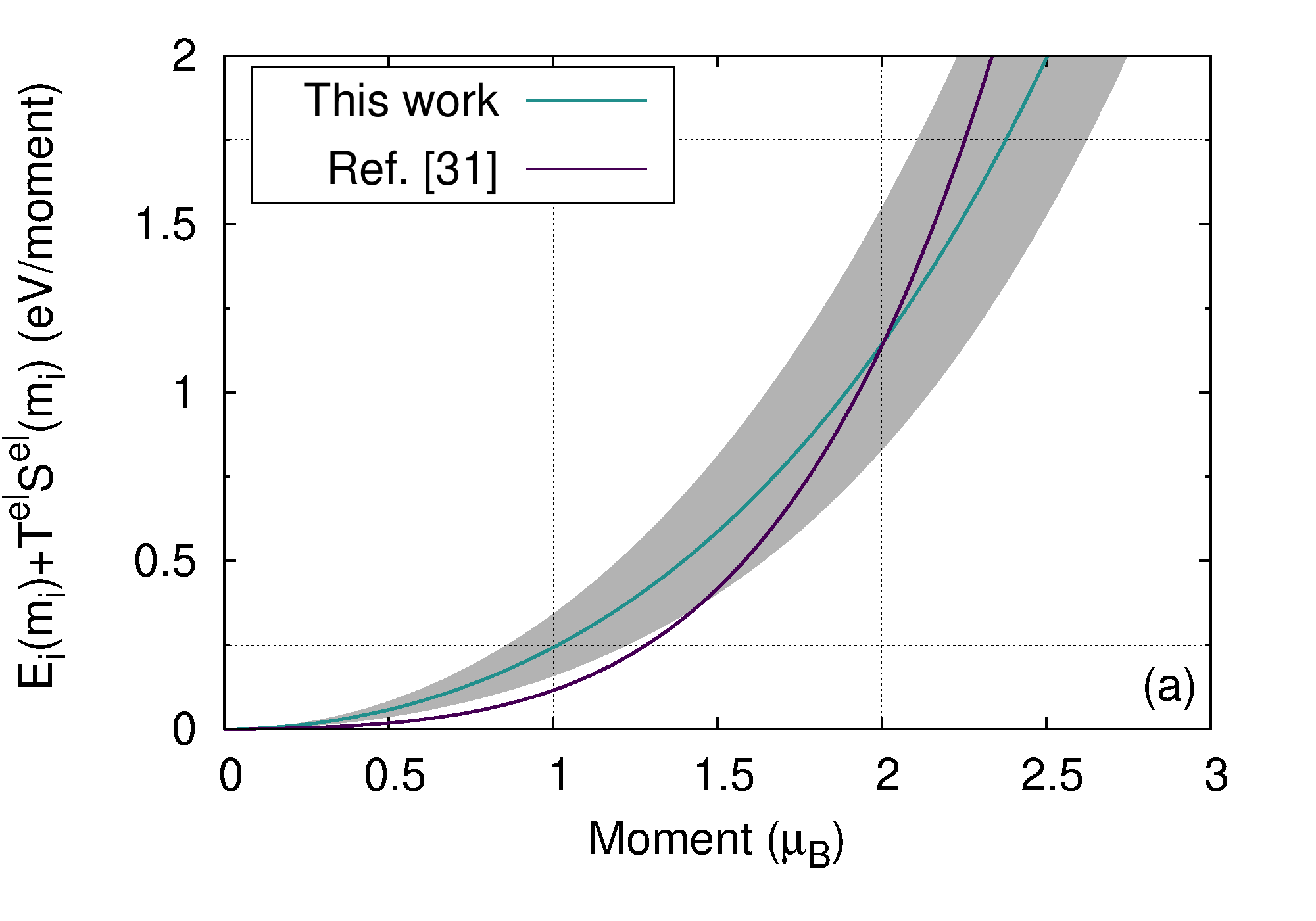}}
\subfigure{ \label{FigAvMFeHTHP}\includegraphics[width=0.5\textwidth]{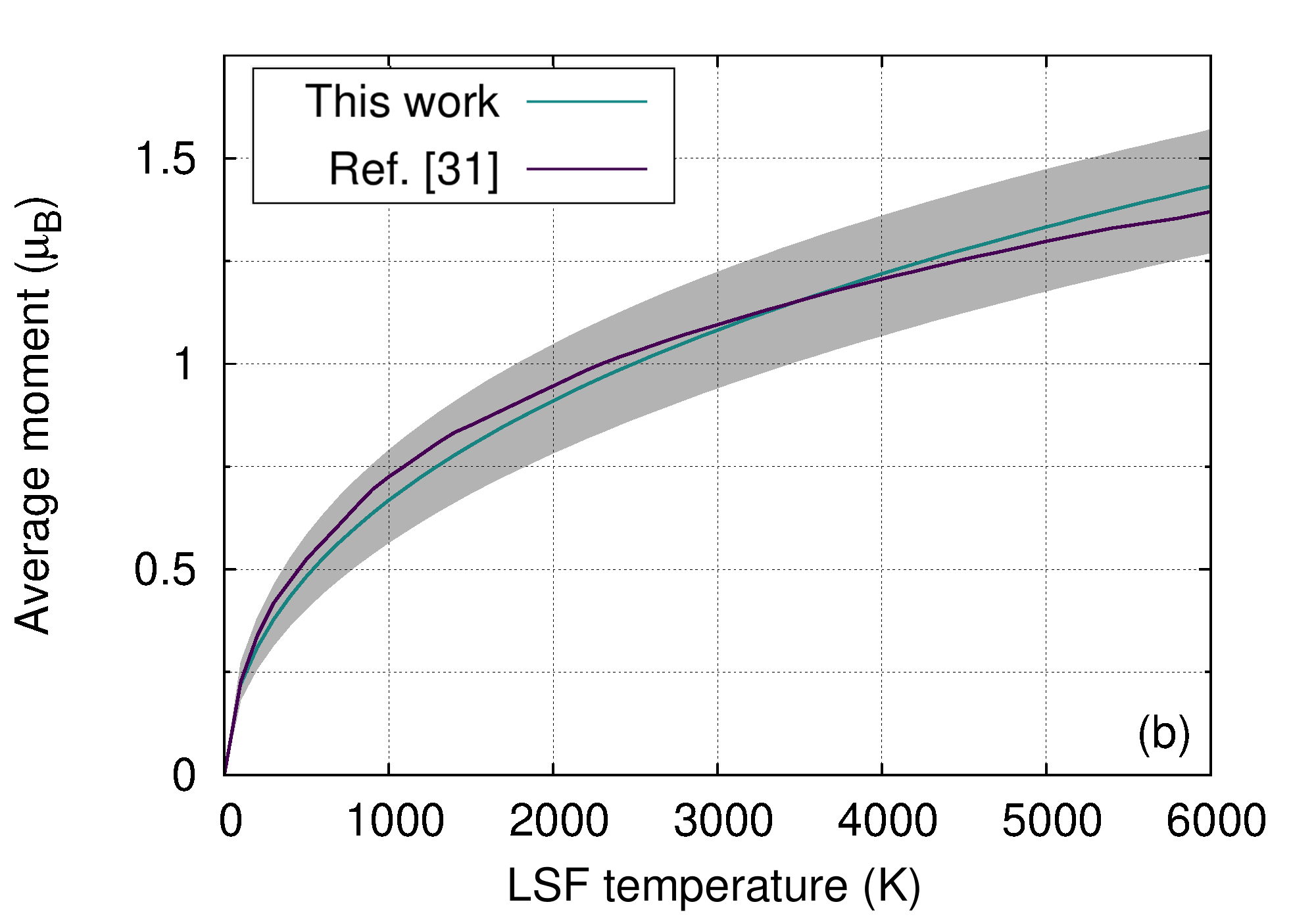}}
\caption{(a) Average landscapes and (b) average moment as a function of LSF temperature for bcc Fe at Earth's inner core conditions from the present work (cyan) compared to the one shown by Ruban $et$ $al.$ in Ref [\onlinecite{RubanEarthCore}]. The gray area in both figures indicates the spread due to unique local environments for each moment. $T^{\textrm{el}}$, the electronic temperature, is set to 6000 K. \label{FigLandscapesAvMFeHTHP}}
\end{figure}

We pass now to the investigation of the effect of LSF in the case of bcc Fe at Earth's inner core conditions, i.e. $T\approx 6000$ K and $p\approx$ 300 GPa. 
It has been speculated \cite{RubanEarthCore,PourovskiiEarthCore} that magnetic effects might play an important role also at these extreme conditions, where local moments on the atoms would rise purely from LSF. 
At these pressures, indeed, all moments die in an unconstrained DFT calculation, indicating a transition to itinerant moment behavior. 
Since the atomic positions in the present calculations are a snapshot of a NM MD simulation (54 atoms in the supercell, $V=7.0$ \AA$^3$/atom) from Ref. [\onlinecite{SergeiFeHTHP}], we start the procedure considering a NM background, i.e., fixing the size of all moments other than the one under investigation to zero.
The NM background is kept only for the first iteration. Convergence is reached after three iterations also in this case.
We employ the three-dimensional PSM, motivated by the purely itinerant nature of the magnetic moments at these conditions. 
The electronic free energy at $T=6000$ K is employed for the energy landscapes, since the purely electronic degrees of freedom are supposed to be faster than the magnetic ones; in addition, employment of electronic energy or free energy gives differences only at a quantitative level in this case.

In Fig. \ref{FigLandscapesFeHTHP} the average landscape obtained in the present work is compared to the landscape calculated by by Ruban $et$ $al.$ \cite{RubanEarthCore}, which was calculated on ideal lattice positions.
The colored area in Fig. \ref{FigLandscapesFeHTHP} indicates the spread in landscapes due to atomic vibrations. 
The average landscape from the present work deviates from the one from Ref. [\onlinecite{RubanEarthCore}], however no appreciable difference is found in the average moment as a function of temperature (Fig. \ref{FigAvMFeHTHP}), with the two curves almost overlapping along the whole temperature range. 

\begin{figure}[t!]
\begin{center}
\includegraphics[width=0.5\textwidth]{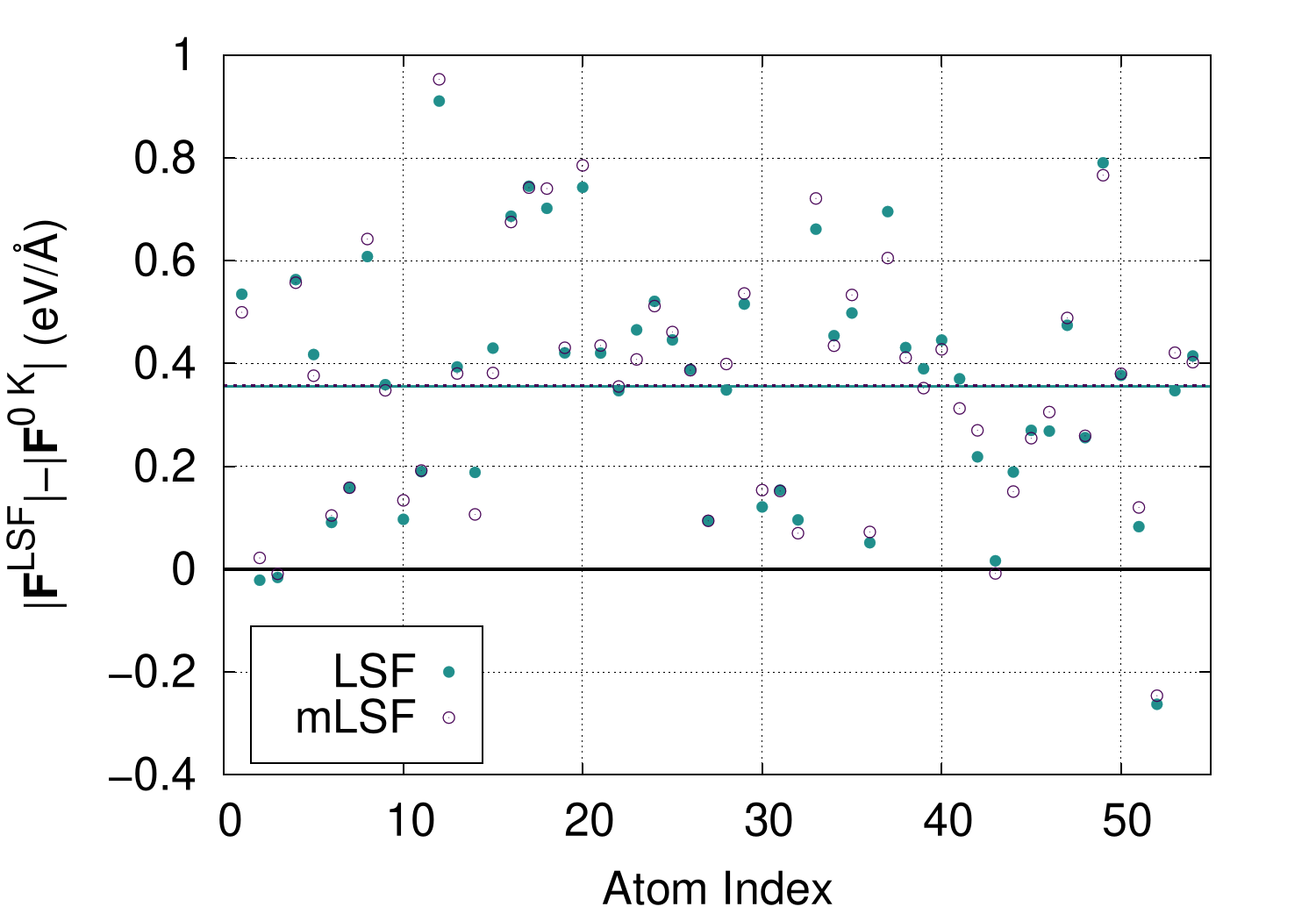}
\caption{Difference in forces modulus between inclusion of LSF and not, both considering local environments effects (solid points) and including LSF only on a mean level (empty points). The value of the moment at 6000 K on a mean level is taken from Ref [\onlinecite{RubanEarthCore}]. The average difference in forces is also depicted by horizontal lines, and the two lines for the different LSF schemes result superposed on each other. \label{FigForcesFeHTHP}}
\end{center}
\end{figure}

It is interesting to estimate the effect on forces deriving from the employment of an average value of magnetic moment size for every atom at 6000 K (taken from Ref. [\onlinecite{RubanEarthCore}]) as compared to full consideration of local environment effects. In Fig. \ref{FigForcesFeHTHP} we therefore compare the difference in forces between the NM calculation and the calculations with LSF, both with consideration of local environment effects and only on mean level. 
The difference from the NM run is quite important, with values as large as $\approx 1$ eV/\AA; however, the two LSF schemes give almost identical results, showing a small effect of individual local environment on the LSF moments and motivating in this case the use of the mean value for future investigations. 

This difference results also in a difference in total pressure in the cell. 
As previously mentioned, the volume used in the present investigation is $V=7.0$ \AA$^3$/atom; for this volume, we obtain a pressure of $\approx 300$ GPa for the NM case, 320 GPa with inclusion of LSF on an average level, and 330 GPa with the present self-consistent calculation of LSF moments.

\begin{figure}[t!]
\begin{center}
\includegraphics[width=0.5\textwidth]{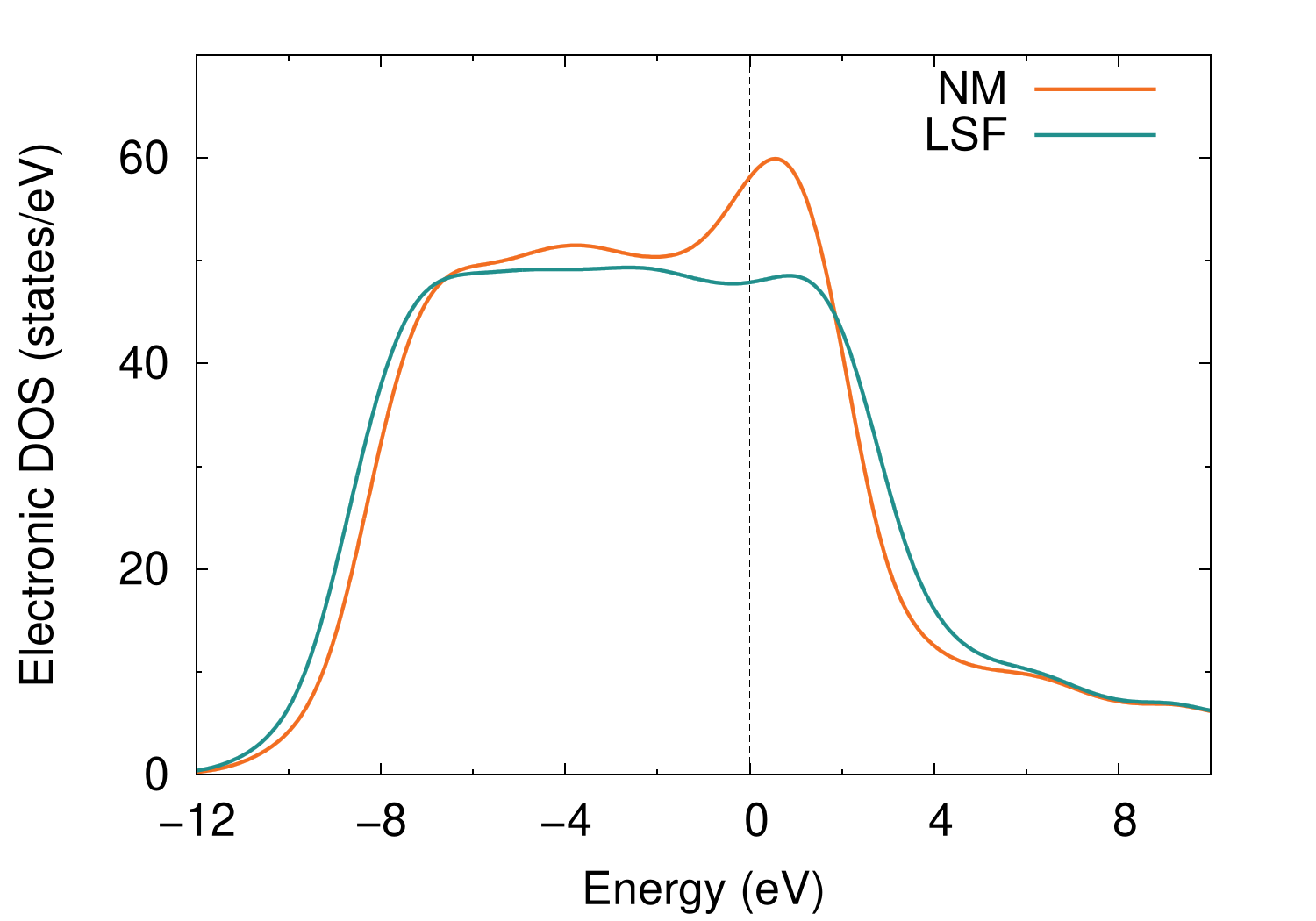}
\caption{Electronic DOS of bcc Fe at Earth's core conditions with and without LSF. Inclusion of LSF induces the complete disappearance of the peak around the Fermi level. \label{FigDOSFeHTHP}}
\end{center}
\end{figure}

Finally, the total electronic DOS of bcc Fe at Earth's core conditions including vibrations with and without LSF is shown in Fig. \ref{FigDOSFeHTHP}. 
We observe a well known peak in the DOS near the Fermi level \cite{PourovskiiEarthCore} for the NM calculation, which is here smooth due to the vibrational disorder in the lattice and the strong electron smearing; the inclusion of LSF leads to its complete disappearance.
This effect is due to the fact that an increased atomic magnetization is achieved by splitting further apart spin-up and spin-down states, inducing a reduction in the DOS at the Fermi level. 
A careful inspection of the consequences of this result is needed and beyond the scope of the present investigation.

\subsection{Liquid Fe at ambient and high pressure} \label{SectionResultsLFe}

The last examples of the power of the present scheme are concerned with liquid Fe at ambient and high ($\approx 200 \;$ GPa) pressure.

\begin{figure}
\begin{center}
\includegraphics[width=0.5\textwidth]{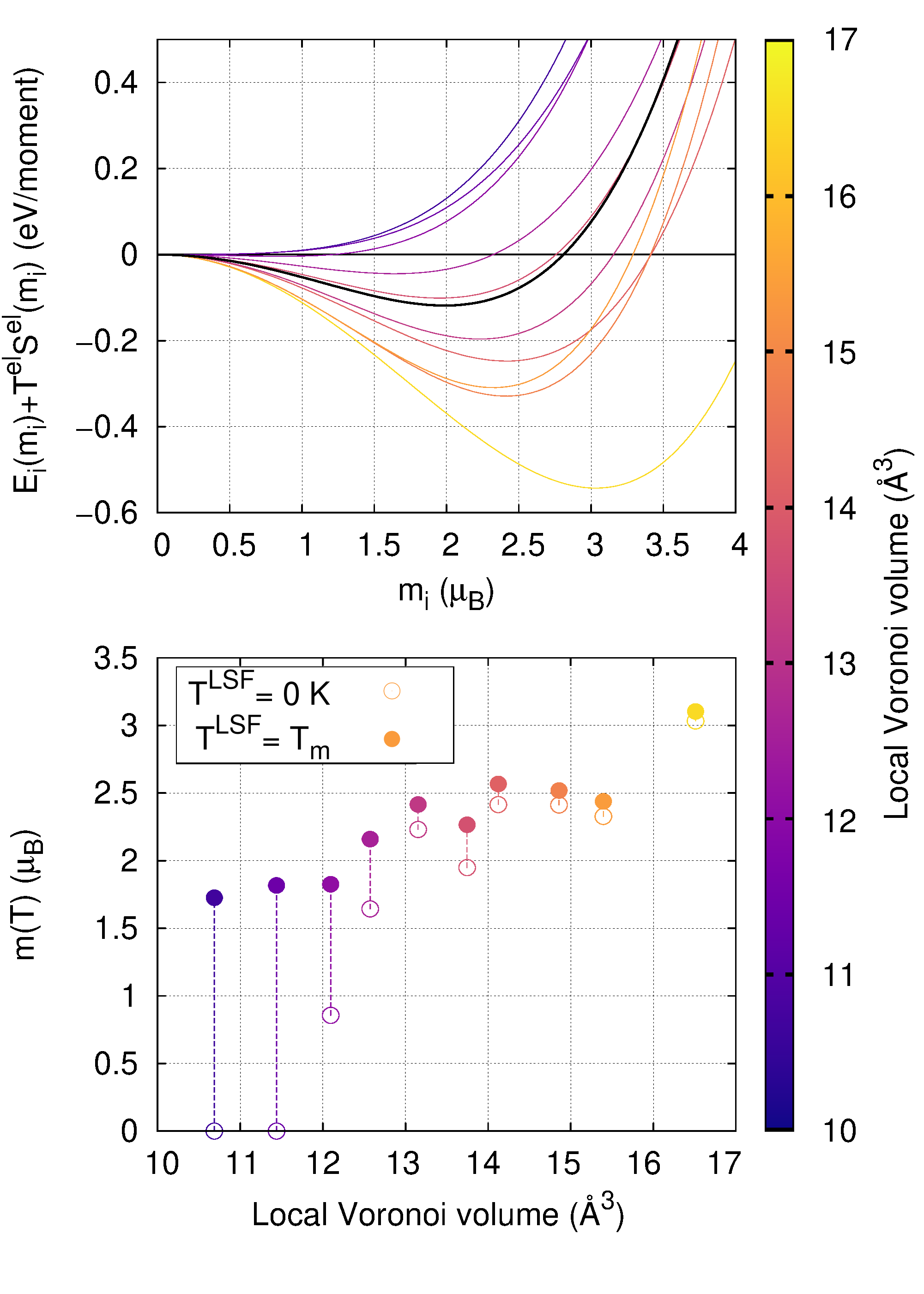}
\caption{(Top) Free energy landscapes calculated for selected atoms in liquid Fe around the melting point, where in the free energy the electronic entropy is included ($T^{\textrm{el}}=T_m$). (Bottom) Moments with LSF temperature of 0 K (empty symbols) and melting temperature (filled symbols) for liquid Fe as a function of Voronoi volume. \label{FigLandscapesMomentsLFe}}
\end{center}
\end{figure}

Starting from the ambient pressure case, where the temperature employed is about the melting point, the atomic configuration is obtained as explained in Sec. \ref{SectionComputationalDetails}. For what concerns the magnetic configuration at this stage of the analysis, noncollinear magnetic moments with random directions were assigned to each atom.
Since the DFT calculations with noncollinear magnetic moments constrained only in direction does not converge, in order to get an initial guess of the size of the moments we calculate the energy landscape for ten atoms in a NM background.
From the following iteration, we assign magnetic moments to every atom assuming a dependence on the Voronoi volumes.
The choice of the ten atoms employed in the analysis is based on their Voronoi volumes, so that we are able to span the whole volume range available in the selected snapshot, which is of course larger than in the solid state case.
In the energy landscape we include also the electronic entropy contribution as in Sec. \ref{SectionResultsbccFeHTHP}.
We employ $\textrm{PSM}=m^2$ and perform three iterations to get the converged moments at $T=T_m$.

In Fig. \ref{FigLandscapesMomentsLFe} the energy landscapes and the magnetic moments with and without LSF are shown.
The landscapes show a larger variety as compared to the solid state counterpart because of the high degree of disorder in the liquid phase. 
LSF tend to increase drastically the size of small moments (corresponding to small Voronoi volumes), whereas they leave roughly unchanged large moments.
In particular, for the atom with very large Voronoi volume ($\approx 17 $\AA$^3$) we get an associated magnetic moment of size $\approx 3 \mu_B$, independently of inclusion of LSF or not, larger than any moment in bcc Fe.

Local magnetic moments are known to be present in liquid Fe from neutron scattering experiments \cite{LFeTmExp}, and they are found to be about 1.2 $\mu_B$. 
This value is much smaller than the effective magnetic moment of 4.4 $\mu_B$ obtained from the measurement of the magnetic susceptibility \cite{LFeExpMagSusc}. 
The discrepancy is attributed in Ref. [\onlinecite{LFeTmExp}] to the role played by conduction electrons in magnetic susceptibility measurements, which for more itinerant systems like fcc Fe and liquid Fe is assumed to be important. 
From our analysis of energy landscapes, we obtain an average magnetic moment of 2.24 and 1.72 $\mu_B$ with and without inclusion of LSF, respectively. 
The latter value is obtained considering the size of the moments at the minimum of the energy landscapes.
The average is weighted with the number of atoms with a Voronoi volume similar to the atoms for which the calculation of the magnetic moment is explicitly performed, so that moment sizes corresponding to frequent Voronoi volumes weigh more than moment sizes with more rare ones.
Our results are considerably larger than the estimate from neutron scattering data both with and without inclusion of LSF. Nonetheless, in Ref. [\onlinecite{LFeTmExp}] a value of the magnetic moment of $0.5-0.8 \mu_B$ for fcc Fe obtained from neutron scattering in a previous work \cite{ExpMomfccFe} was reported; from DFT calculations, this is found to be $\gtrsim 2 \mu_B$ (see, e.g., Ref. [\onlinecite{LSFVitos}]) also without LSF.
Therefore, the experimental estimate might be underestimating the size of the magnetic moments in liquid Fe. 
In addition, in this work we have taken into account only one atomic-magnetic configuration, and the inherently disordered nature of liquids requires a much broader statistical sampling to obtain reliable results, which could be then compared to experiments or higher level theories.
This is clearly beyond the scope of the present analysis.

\begin{figure}[t!]
\subfigure{ \label{FigLandscapesLFeHTHP}\includegraphics[width=0.5\textwidth]{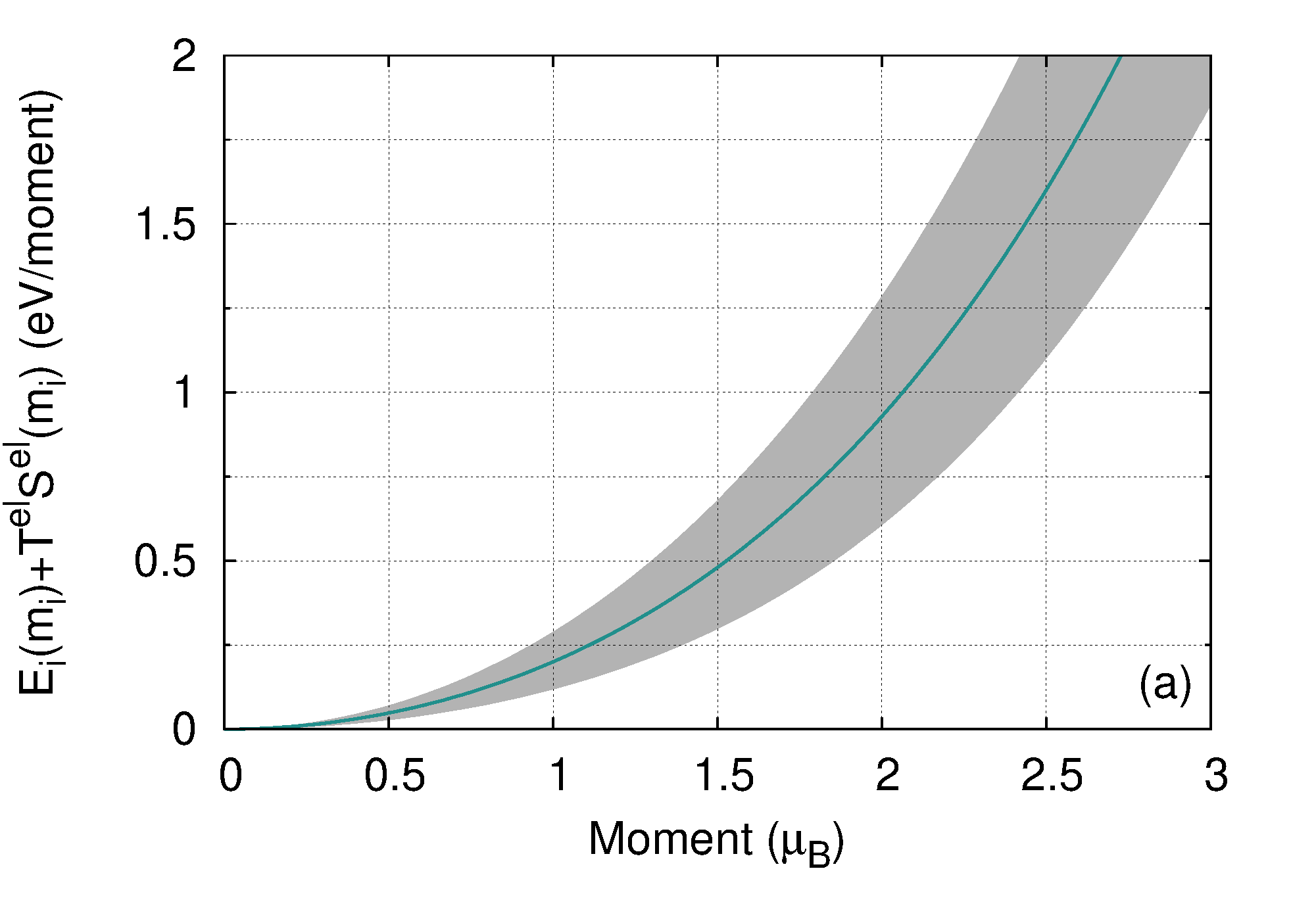}}
\subfigure{ \label{FigAvMLFeHTHP}\includegraphics[width=0.5\textwidth]{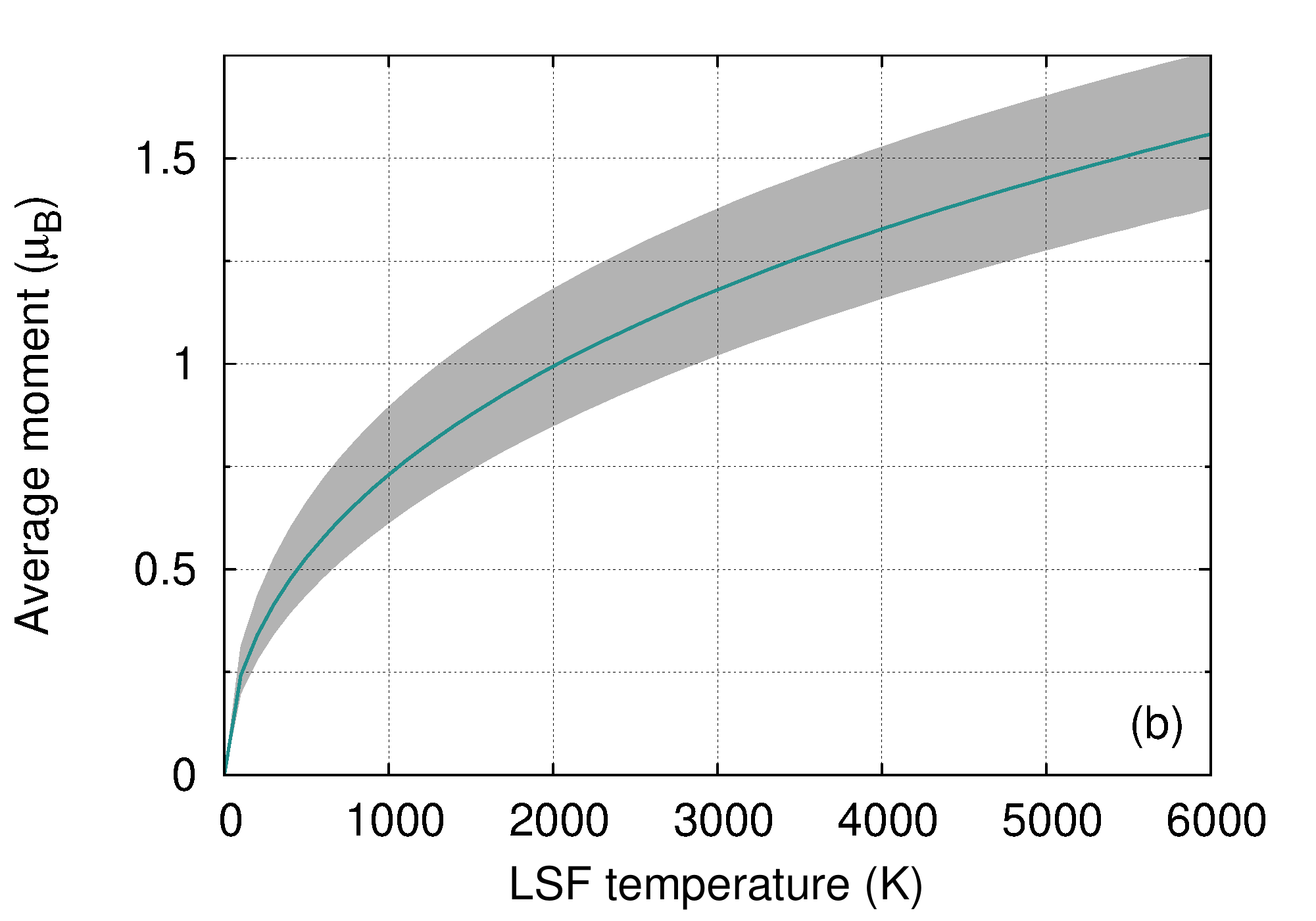}}
\caption{(a) Average landscapes and (b) average moment as a function of LSF temperature for liquid Fe at Earth's outer core conditions. The grey area in both figures corresponds to the spread due to unique local environments for each moment. As in Fig. \ref{FigLandscapesFeHTHP}, the electronic temperature is set to 6000 K. \label{FigLandscapesAvMLFeHTHP}}
\end{figure}

For what concerns the case at Earth's outer core conditions, the moments die off in a regular DFT calculation, as happens in the solid at high pressure (see Sec. \ref{SectionResultsbccFeHTHP}). 
In Fig. \ref{FigLandscapesAvMLFeHTHP} the average free energy landscape and average moment as a function of temperature are shown for this case, where $\textrm{PSM}=m^2$ is employed. 
As in the previous cases, the landscape is considered with inclusion of the electronic entropic contribution at $T=6000$ K.
We observe a less steep free energy landscape as compared to the solid state case (compare with Fig. \ref{FigLandscapesFeHTHP}), which results in larger moments at high temperature.
Of  course, this has also an effect on the pressure in the system, going from $\approx 190$ GPa in the NM case, to $\approx 220$ GPa in the case with LSF; the increase in pressure due to LSF is similar to the solid state case.
These results contradict the framework in which Ref. [\onlinecite{LFeHTHPtheoPRL}] was carried out, where a transition from paramagnetic to diamagnetic in liquid Fe was observed by collinear spin-polarized AIMD above 50 GPa. In that work, any contribution from LSF was neglected, therefore the suggested transition is not supported by enough evidences. In the present work, we find that magnetic fluctuations could play a role even at larger temperatures and pressures than the ones considered in Ref. [\onlinecite{LFeHTHPtheoPRL}], including conditions of Earth's outer liquid core.

\section{Conclusions} \label{Conclusions}

In this work we have developed a supercell approach for the derivation of the finite temperature size of magnetic moments in magnetic materials, based on the semiclassical theory of longitudinal spin fluctuations, which enables investigation of this degree of freedom in presence of structural, vibrational, magnetic, and even liquid disorder in a self-consistent way.
The work is carried out on the level of constrained noncollinear DFT calculations.

We find for bcc Fe at $T\approx T_C$ that lattice vibrations affect the on-site energy landscape as a function of moment size, leading to some more itinerant moments as compared to the usual localized behavior; increasing degree of disorder in the system makes the average energy landscape shallower. 
The choice of phase space measure is critical in this system, since for three-dimensional PSM the average magnetic moment increases as compared to the ``0 K'' size, whereas for the monodimensional PSM this quantity decreases.
LSF affect also forces intensity and direction.

In the case of bcc Fe at Earth's inner core conditions, we find that LSF induce a substantial magnetic moment on the atom which strongly affects the forces in the system as compared to nonmagnetic calculations. 
The present results are in good agreement with Ref [\onlinecite{RubanEarthCore}], suggesting that in this system LSF can be included even at an average level, without consideration of local environment effects.

We perform the same analysis also for liquid Fe at ambient pressure and temperature about the melting point, and at high pressure and temperature, conditions expected at Earth's outer core. 
In the former case, we find local moments also in ``0 K'' calculations, which are then increased when LSF are included in the analysis. 
In the latter, similarly to the solid state case at Earth's inner core conditions, LSF stabilize local magnetic moments, which can be very important for geodynamics of Earth's outer liquid core.

\section*{Acknowledgments}
This research was carried out using computational resources provided by the National Supercomputer Centre (NSC) in Link\"oping (Sigma supercomputer) and the Swedish National Infrastructure for Computing (SNIC): Tetralith Cluster located at NSC.
B.A. acknowledges financial support by the Swedish Research Council (VR) through the International Career Grant No. 2014-6336 and the Grant No. 2019-05403, by Marie Sklodowska Curie Actions, Cofund, Project INCA 600398, from the Swedish Government Strategic Research Area in Materials Science on Functional Materials at Link\"oping University (Faculty Grant SFOMatLiU No 2009 00971), from the Knut and Alice Wallenberg Foundation (Wallenberg Scholar Grant No. KAW-2018.0194), as well as support from the Swedish Foundation for Strategic Research through the Future Research Leaders 6 program, FFL 15-0290. 
Andrei Ruban is gratefully acknowledged for useful discussions. DG thanks Johan Klarbring for interesting discussions.

\end{document}